\documentclass[aps,prl,twocolumn,groupedaddress]{revtex4}
\usepackage[pdftex]{graphicx,epsfig}

\begin{document}

\title{Internal network dynamics prolong a losing battle} 

\author
{Zhenyuan Zhao$^{1}$, Juan Camilo Bohorquez$^{1,2,3}$, Alex Dixon$^4$ \& Neil F. Johnson$^{1}$}
\affiliation{$^{1}$Physics Department, University of Miami, FL 33126, U.S.A.\\
$^{2}$CEIBA Center for Complexity and Industrial Engineering, Universidad de Los Andes, Bogota, Colombia\\
$^{3}$Human Conflict Analysis Center (CERAC), Bogota, Colombia\\
$^{4}$Physics Department, Cambridge University, Cambridge CB3 0HE, U.K.}

\date{\today}

\begin{abstract} 
Fights-to-the-death occur in many natural, medical and commercial settings. Standard mass action theory and conventional wisdom imply that the minority (i.e. smaller) group's survival time  decreases as its relative initial size decreases, in the absence of replenishment. Here we show that the {\em opposite} actually happens, if the minority group features internal network dynamics. Our analytic theory provides a unified quantitative explanation for a range of previously unexplained data, and predicts how losing battles in a medical or social context might be extended or shortened using third-party intervention.

\end{abstract}

\maketitle

Wars of attrition are widespread, e.g. fights among 
ants\cite{nicola}, chimpanzees\cite{mike2,mike1} or 
birds\cite{Radford}, a War of the Worms on the 
Internet\cite{warworms}, commercial wars between companies or products\cite{yano}, human warfare\cite{claus,lanchester,epstein,Panchev}, immunological battles against 
disease\cite{immune,maini,wu1}, irreversible biochemical reactions\cite{kepes}, and even electron-hole recombination in semiconductors\cite{quiroga}.  Much fascinating work exists on predator-prey problems\cite{McKane}, however little has appeared concerning the duration $T$ of a predator-predator `fight-to-the-death'. The duration $T$ would traditionally be calculated invoking a mass-action law for the irreversible reaction $A+B\rightarrow Z$, where $Z$ is neither $A$ nor $B$.
Standard mass action theories\cite{lanchester,mangel} and conventional military thinking\cite{claus} both suggest that the minority (i.e. smaller) group's survival time $T$ decreases as its relative initial size decreases. 

In this paper, we calculate analytically the duration $T$ of a war of attrition, uncovering the surprising result that the adoption of internal network dynamics greatly prolongs the minority group's survival time. By contrast, $T$ is remarkably insensitive to the internal dynamics adopted by the majority population. Our analytic theory provides a unified, quantitative explanation for a range of previously unexplained data. It also predicts how the duration of a losing battle can be manipulated using modest third-party intervention and without the need to directly fight
either predator. In addition to its potential applications, our work provides a novel demonstration of the breakdown of standard mass-action differential equations when describing a system with dynamical internal heterogeneity.   

The mass-action equations which are routinely applied to conflicts take one of two forms\cite{lanchester,nicola,mike2,mike1,yano,epstein,Panchev,
wu1}: (1) 
$d N_A(t)/dt=-a
N_A(t) N_B(t)$ and $d N_B(t)/dt=-b N_A(t) N_B(t)$, which is called Lanchester's undirected fire model; or (2) $ d N_A(t)/dt=-a' N_B(t)$ and $d N_B(t)/dt=-b'
N_A(t)$, which is called Lanchester's directed fire model\cite{lanchester,nicola,mike2,mike1,yano,epstein,Panchev,
wu1}. Here $N_A(t)$ and $N_B(t)$ are the population sizes at time $t$  and $a$, $b$, $a'$, $b'$ are constants. For simplicity, we focus on simple attrition where no appreciable  replenishment of $A$ or $B$ arises over the timescale of the entire process, and we adopt the familiar language of human conflict -- however these simplifications can be generalized according to application area\cite{wu1}. Indeed, 
generalized versions of these mass-action models lie at the heart of modern mathematical biology\cite{epstein,mangel}. The blue curves in Fig. 1 show these two mass-action models' predictions for the time $T$ taken to eliminate the smaller population, as a function of the asymmetry $x$ between peak (i.e. initial) population size: $x=|N_A(0)-N_B(0)|/[N_A(0)+N_B(0)]$, hence $x\sim 0$ when the peak (i.e. initial) population sizes are similar and $x\sim 1$ when very different. The total number is fixed at $N=N_A(0)+N_B(0)$. Since $N\gg 1$, we can take the end-point for the undirected fire model to correspond to reducing the smaller population to one instead of zero, thereby avoiding problems with a continuum description of $N_A(t)$ and $N_B(t)$ near zero. We set $a=b$ and $a'=b'$ to avoid any implicit bias.

\begin{figure}
\includegraphics[width=0.53\textwidth]{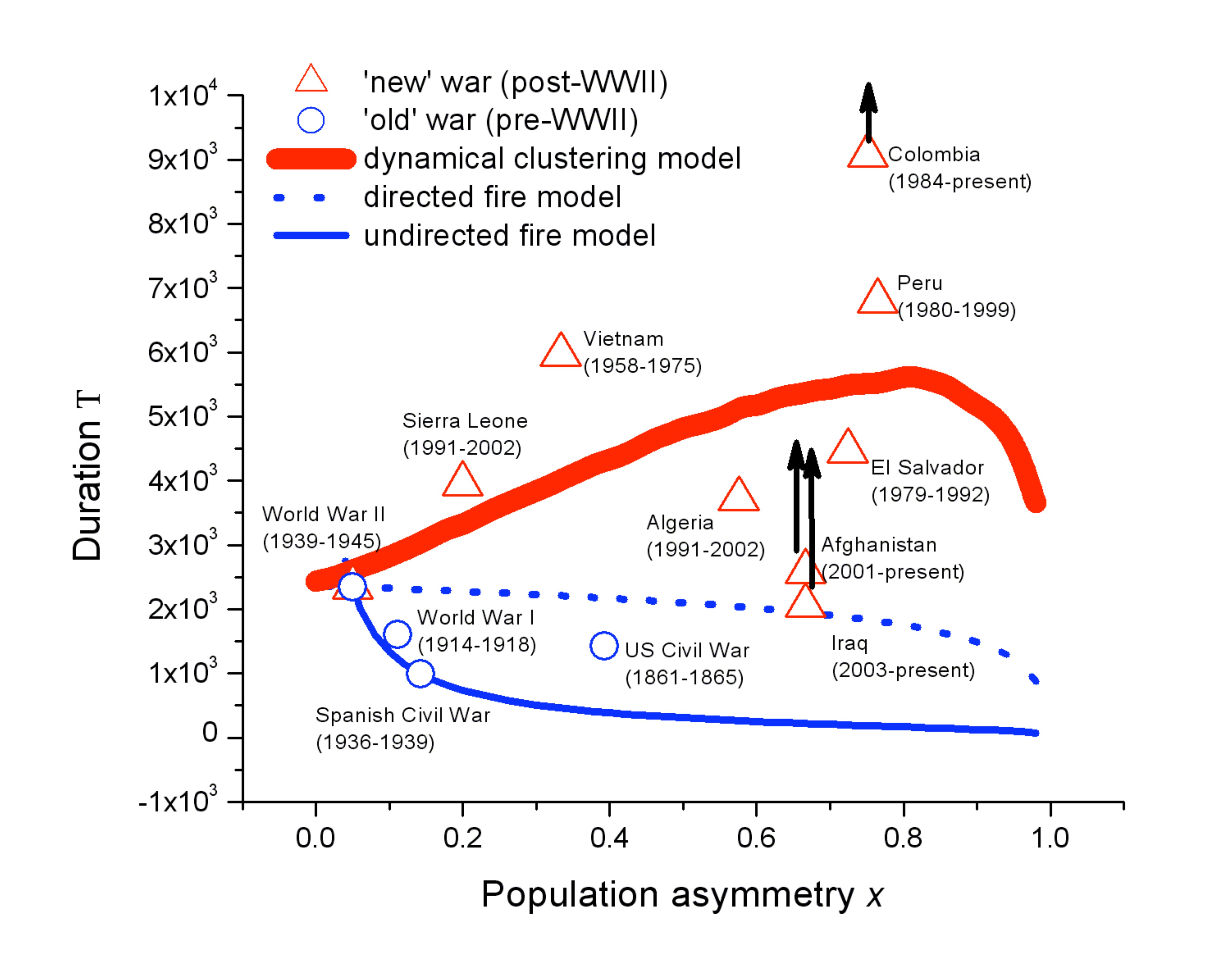}
\caption{(Color online) Vertical axis shows duration $T$ (or equivalently, the extinction time or survival time of the minority group)
as a function of  the asymmetry $x$ between the two opposing forces. Blue circles are `old' wars, red
triangles are `new' wars\cite{kaldor,robb}. Data are up to the end of 2008, hence the final datapoints for the three ongoing wars will lie {\em above} the positions shown, as indicated by arrows. Lower two blue lines are the mass-action (i.e. Lanchester) model results.
Upper red line is our theoretical curve (i.e. Eq. (3)) with $\nu_{A}=\nu_{B}=0.7$.  Changing  $\nu_{A}$ and $\nu_{B}$ changes the height of the theoretical peak, but leaves the qualitative features unchanged.}
\end{figure}

In the case that $B$ (or $A$) features internal network dynamics, our model represents a generalization of Ref. \cite{ez}. The resulting population, which evolves continually in time, features (i) temporary communities (i.e. clusters) which may form, grow and breakup over time, (ii) a fluctuating number of such communities and community membership, (iii) no permanent central leadership and no permanent hierarchical structure, and (iv) long-ranged interactions. While explicit spatial generalizations are possible, this model already mimics the fluid, non-spatial community structure expected of traders in a global financial market\cite{ez,neilfin}, or insurgent cells in a modern guerilla war in which communication can take place over any distance\cite{kaldor,robb,njohnson}. 
At time $t$, populations $A$ and $B$ have $n^A_{s}(t)$ and $n^B_{s}(t)$ clusters
of size $s$, such that $\sum sn^A_{s}(t)=N_{A}(t)$ and $\sum sn^B_{s}(t)=N_{B}(t)$. A cluster is picked
randomly with a probability proportional to its
size, mimicking the notion that at any moment in time any
randomly chosen individual can
initiate an event. Figure 2(c) summarizes the corresponding events. This cluster fragments with
a probability dependent on its identity, i.e. $\nu_{A}$ or
$\nu_{B}$, mimicking the idea of clusters disbanding if they sense danger. If the cluster does not fragment, a second cluster is
selected from the total population. If the two clusters are both $A$ or both $B$, they coalesce. This `coalescence' can simply mean that the two groups act in a coordinated way, not necessarily that they are physically joined. If they are of different types, they interact (i.e. fight). During this interaction, the smaller cluster is destroyed
and the larger cluster is reduced in size by an amount equal to the
smaller cluster's size -- if they are the same size, 
both clusters are destroyed. Hence both
populations lose the same number of objects which are then removed, however we stress that our main results are unchanged if the clusters
suffer only finite losses or receive a limited number of new recruits. 
The probability $Q_{AB}$ that
any $A$ cluster is selected and interacts with a $B$ cluster is the sum over all $s$ of the probability for an $A$ cluster of size \(s\) to interact with any $B$ cluster, which gives 
$(1-\nu_{A}) {N_{A}(t) N_{B}(t)}/[N_{A}(t)+N_{B}(t)]^{2}$. The
probability $Q_{BA}$ is a similar expression,
with $\nu_{A}$ replaced by $\nu_{B}$. After an interaction, $A$ and $B$ are reduced in number by the size of the smallest interacting cluster. The average interaction size $c$ is well approximated by unity, with a small linear correction term $0.2
{N_B(0)}(1-\nu_{A})(1-\nu_{B})/{N_A(0)}$. 
The
populations after $i$ interactions become $N_A(t)=N_{A}(0)-ic$, $N_B(t)=N_{B}(0)-ic$. 
Hence the probability for an interaction between $A$ and $B$ clusters after \(i\) previous interactions is
\begin{eqnarray}
Q(i) &=& Q_{AB}\! +\! Q_{BA} \\
&=& \frac{{(N_A(0) - ic)(N_B(0) - ic)}}{{(N_A(0) + N_B(0) - 2ic)^2}}(2 - \nu_A -\nu_B)  \nonumber
\end{eqnarray}
To reduce $N_A(t)$ and $N_B(t)$ by \(c\) takes \(1/Q(i)\) timesteps on average. The total time to reduce one population to zero is the sum of the timesteps required for each interaction, until the population is wiped out.
Supposing $B$ is the smaller population, it will require \(N_B(0)/c\) interactions to destroy it, hence the final interaction will happen after \(N_B(0)/c - 1\) previous interactions. The total time to reduce the smaller population $B$ to zero is therefore
\begin{eqnarray}
T &=& \sum\limits_{i = 0}^{\frac{{N_B(0)}}{c} - 1} {\frac{1}{{Q(i)}}}  \\
&=& \sum\limits_{i = 0}^{\frac{{N_B(0) }}{c} - 1} {\frac{{(N_A(0)  + N_B(0)  - 2ic)^2}}{{(N_A(0)  - ic)(N_B(0)  - ic)(2 - \nu _A  - \nu _B )}}\ \ .  \nonumber} 
\end{eqnarray}
\noindent Using the mathematical relationships \(\sum_{1}^{n}{\frac{1}{i}} = \gamma + \psi_0(n+1)\), where \(\gamma\) is the Euler-Mascheroni constant and \(\psi_0\) is the digamma function, and \(\sum_{a+1}^{n} = \sum_{1}^{n} - \sum_{1}^{a}\), we obtain the final expression for the conflict's duration:
\begin{eqnarray}
\small
\small
\label{eqn:extinction}
T &=& \frac{N_A(0)-N_B(0)}{c(2-\nu_{A}-\nu_{B})}
\Big[\frac{4N_B(0)}{N_A(0)-N_B(0)}\nonumber \\
& & +\big[\gamma+\psi_{0}\Big(\frac{N_B(0)}{c}+1\Big)\big]\\ 
& &-\big[\psi_{0}\Big(\frac{N_A(0)}{c}+1\Big)-\psi_{0}\Big(\frac{N_A(0)-N_B(0)}{c}+1\Big)\big]
\Big]\ \ . \nonumber 
\end{eqnarray}
If $A$ is the smaller population, the equation has an identical form but with $A$ and $B$ interchanged.

The red curve in Fig. 1 shows the analytic curve $T$ from Eq. (3). The peak at $x\sim 0.8$ is robust to many model variants. Its origin can be understood as follows: When $x\sim 0$, clusters of $A$ and $B$ are abundant and have a reasonably large average size. Encounters between $A$ and $B$ clusters are frequent and the attrition per encounter is high, hence $T$ is small. 
As $x$ increases, with $A$ being the larger force, an encounter between an $A$ and $B$ cluster is increasingly likely to wipe out the $B$ cluster completely since the $A$ cluster is increasingly likely to be the larger cluster. However the encounter rate is decreasing, and the overall effect is that $T$ increases. For $x\rightarrow 1$, it still takes a long time to find a $B$ cluster however there are now very few to find, hence $T$ now decreases.  Interestingly, the distribution of inter-event times changes from near-exponential away from the peak, to
near-power-law around the peak. 
Figure 2(a) confirms the accuracy of our analytic curve as compared to numerical simulations, the expected symmetry under $A\leftrightarrow B$, and the insensitivity of the main features to the fragmentation and coalescence probabilities. We note that if the conflict were to end after a given fraction of the initial population is removed, the same qualitative results still emerge since the theory is essentially invariant under an overall change of scale. Also, members of $A$ and $B$ need not actually be destroyed during an encounter -- they just have to disappear from the fighting force. Hence the attrition  of $N_A(t)$ and $N_B(t)$ in Fig. 1 is also interpretable as a loss of strength or will-power, as opposed to actual casualties.

What if $A$ or $B$ adopt different internal dynamics? The black curve in Fig. 2(b) reproduces the analytic result from Fig. 1, while the red-dashed and green curves explore other scenarios. Remarkably, the duration $T$ remains essentially unchanged if the majority group chooses a static internal structure comprising rigid units of any particular size. Instead, the internal network dynamics of the minority group dictate the duration $T$. If the minority group loses its dynamical clustering ability and becomes internally static (i.e. rigid) the duration $T$ decreases significantly, as seen from the black and red-dashed curves for small $N_A(0)$. If both $A$ and $B$ are internally static, $T$ is uniformly small for all $x$. 

Empirical validation of Eq. (3) is possible using published durations for human conflicts. Indeed, an intriguing distinction has recently been hypothesized between old wars in which $A$ and $B$ adopt traditional, fairly rigid, military structures, and new wars in which $B$ (and possibly $A$) adopt more fluid tactics akin to our model\cite{kaldor,robb}. Figure 1 demonstrates that our results offer a quantitative validation of this hypothesis. `Old' wars are blue circles and `new' wars are red triangles, with World War II labelled by both since it is a natural dividing point. Despite the simple generic nature of our model as compared to the complexity and variety of real-world conflicts, Eq. (3) captures the trend for all the `new' wars\cite{kaldor,robb}, suggesting that our proposed internal network dynamics for the minority population represents a key common feature determining their duration. By contrast, the `old' wars are well described by both the green curve of Fig. 2(b) (i.e. rigid armies) and the predictions of conventional mass-action theory (blue curves in Fig. 1)\cite{lanchester,claus}, implying that such internal network dynamics were absent in `old' wars. Additional validation is available in the form of vocal bird rallies in the wild, since bird calls are very long-ranged and birds cluster dynamically\cite{Radford}.
If all possible $N_A(0): N_B(0)$ ratios are assumed equally likely over time, as expected in the wild, the histogram obtained by integrating our theoretical $T$ vs. $x$ curve over all $x$ will be bimodal, with peaks at small and large $T$, which is precisely the reported empirical form\cite{Radford}.

\begin{figure}
\includegraphics[width=0.5\textwidth]{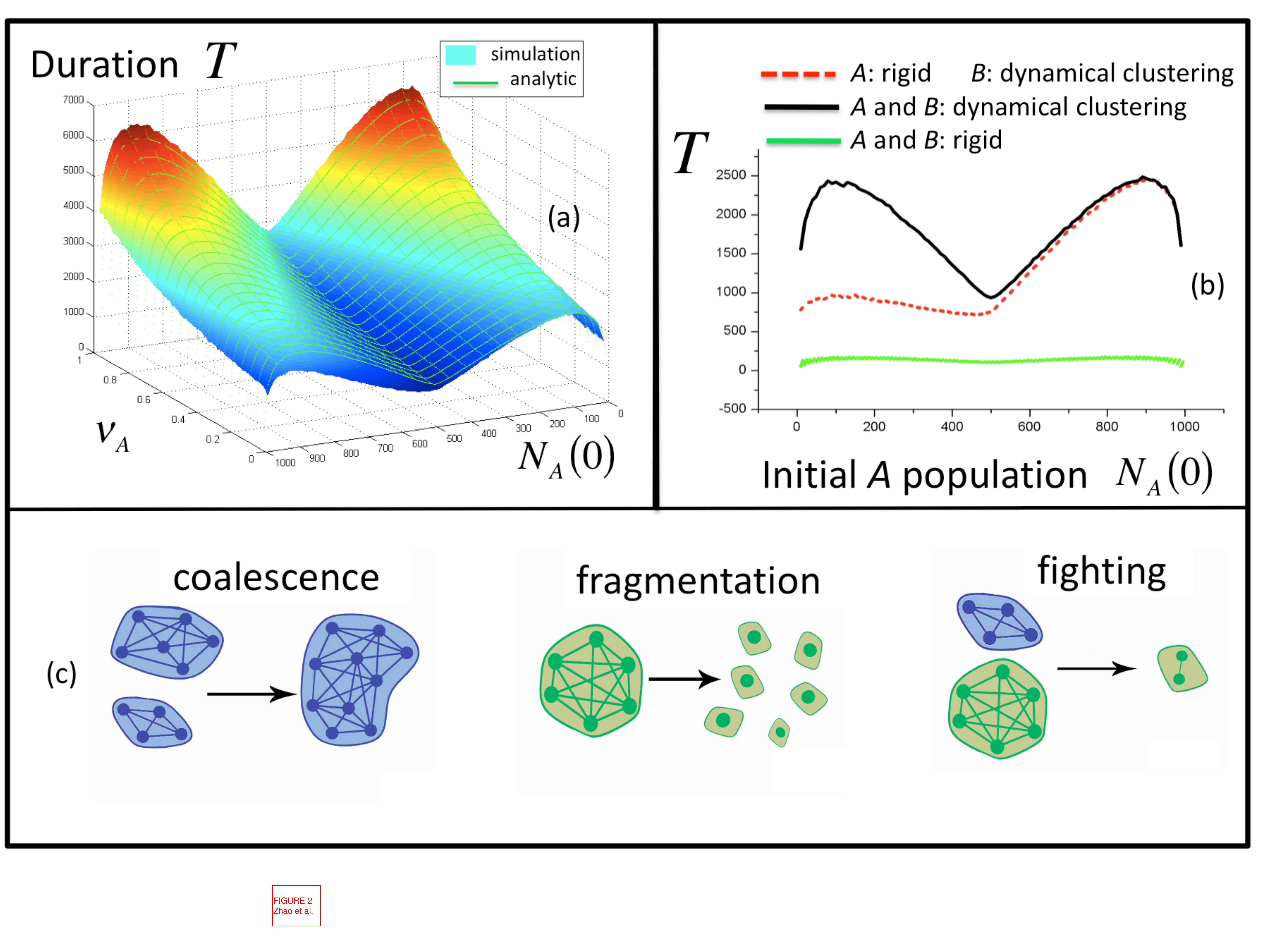}
\caption{\label{fig:extinction3} 
(Color online)  (a) Solid lines are analytic (Eq. (3)) while the surface is a
numerical simulation, as a function of initial $A$ population $N_A(0)$ and fragmentation probability $\nu_A$ ($\nu_{B}=0.3$). Right-hand branch contains the
red curve from Fig. 1. Initial population $N_A(0)+N_B(0)=1000$.
(b) Black curve: same as Fig. 2(a) with $\nu_{A}=\nu_{B}=0.3$. Red dashed curve: $A$ contains rigid units (e.g. size 10) and $B$ is dynamically clustering. Green curve: both $A$ and $B$ comprise rigid units (e.g. size 10).
(c) Events. Nodes represent
individuals, or equivalently some indivisible basic fighting unit. Intra-cluster connections are very strong, while inter-cluster connections are very weak.  Two clusters of same type can coalesce (e.g. $6+4=10$). Individual clusters can fragment (e.g. $6\rightarrow 6\times 1$). Two clusters of opposite type will fight. If a size $6$ cluster fights a size $4$ cluster, a
single cluster of size $6-4=2$ survives.}
\end{figure}

\begin{figure}
\includegraphics[width=0.5\textwidth]{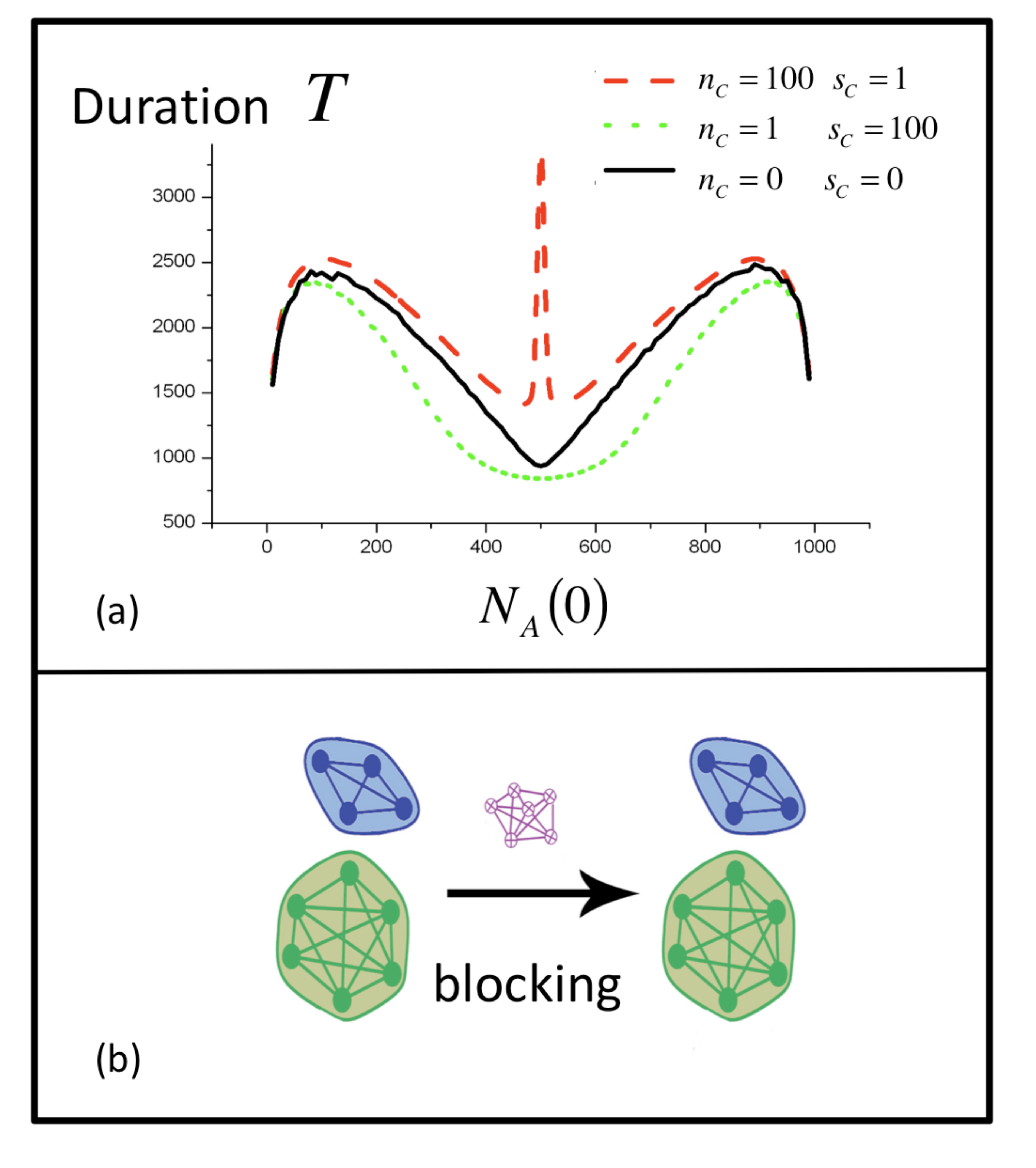}
\caption{\label{fig:variations} (Color online) 
(a) Black curve as in Figs. 1, 2(a) and 2(b) with $A$ and $B$ undergoing
dynamical clustering. $\nu_A=\nu_{B}=0.3$. Red dashed curve: $n_C=100$ third-party groups, each of size $s_C=1$. Green curve: $n_C=1$ third-party group, of size $s_C=100$. 
(b) Third-party blocking event. When both
$A$ and $B$ clusters are smaller than the $C$ cluster, the $A$ and $B$ clusters are neutralized.}
\end{figure}

Figure 3(a) shows that the duration $T$ can be manipulated significantly by adding a third-party 
population $C$ which can block fights (see Fig. 3(b)). For simplicity, we assume here that the  $N_C$ members of $C$ are permanently arranged into $n_C$ groups
each with $s_C$ permanent members. Apart from peacekeepers in human conflict, $C$ could for example be chosen to block specific self-interaction processes in a damaged immune system\cite{immune,kepes}. $A$ and $B$ undergo dynamical clustering as before, except that if a $C$ group is selected and it is bigger or equal to the size of the
$A$ and $B$ clusters, the interaction is blocked and the two $A$ and $B$ clusters are permanently pacified (i.e. neutralized). Figure 3(a) shows that the duration curves undergo a highly non-trivial change.
If $C$ comprises a only a few, large groups (e.g. green dotted curve) then $T$ decreases irrespective
of the asymmetry since $C$ clusters will tend to dominate all encounters. By contrast, if $C$
comprises many small groups (e.g. red dashed curve) $T$ can
be much larger, showing a huge increase around $x\sim 0$ since the $C$ clusters still get picked for encounters but are not large enough to neutralize the parties -- hence their presence simply delays the end of the war.

Many generalizations of our model can be made without affecting the main findings, e.g. $A$ and $B$ can be replenished proportionally by a slow birth or recruitment process. In addition to their theoretical interest, we believe our results may have particular applicability to immunological battles\cite{immune,wu1} where dynamical networks and clusters with long-ranged non-local interactions are possible\cite{kepes}. For example, our results may help shed light on why some latent diseases survive so long against the immune system, and why small but mobile populations of metastic cancer cells can be so difficult to eradicate\cite{maini}.


\begin{thebibliography}{99}
\bibitem{nicola} N. Plowes and E. Adams, {\em Proc. R. Soc. B} {\bf 272}, 1809 (2005).
\bibitem{mike2} M. Mesterton-Gibbons, 
{\em An Introduction to Game-Theoretic Modelling}
(American Mathematical Society, New York, 2000).
\bibitem{mike1} E. Adams and M. Mesterton-Gibbons, {\em Behav. Ecol.} {\bf 14}, 719  (2003).
\bibitem{Radford} A. Radford and M. Du Plessis, \emph{Anim. Behav.} \textbf{68,} 803 (2004).
\bibitem{warworms} S. Tanachaiwiwat and A. Helmy, Technical Report CS 05-859, University of Southern California (2005).
\bibitem{yano} S. Yano, {\em New Lanchester strategy}, (Lanchester, California, 1996).
\bibitem{claus}  C. Von Clausewitz, {\em On War} (Trubner, London, 1873).
\bibitem{lanchester} N. MacKay, \emph{Math. Today}, {\bf 42}, 170 (2006).
\bibitem{epstein} J. Epstein, {\em Nonlinear Dynamics, Mathematical Biology and Social
Sciences} (Addison-Wesley, Reading, 1997).
\bibitem{Panchev} S. Panchev and N. Vitanov, e-print lanl.arxiv.org/abs/0810.3818v1
\bibitem{immune} H. Ploegh, {\em Paradigm Magazine}, Nov. 16, 2005.
\bibitem{maini} R. Gatenby and P. Maini, \emph{Nature} {\bf 421}, 321 (2003).
\bibitem{wu1} G. Wu, and S. Yan, {\em Comp. Clin. Path.} {\bf 11}, 113 (2002); {\em Amer. Jour. of Infect. Diseases}. {\bf 1}, 156, (2005); {\em Comp. Clin. Pathol.} 
{\bf 11}, 178 (2005).
\bibitem{kepes} R. Callard and J. Stark, in {\em Biological Networks} (World Scientific, London, 2007).
\bibitem{quiroga} F. Rodriguez, L. Quiroga, C. Tejedor, M. Martin, L. Vina and R. Andre, {\em Phys. Rev. B} {\bf 78}, 035312 (2008).
\bibitem{McKane} A. McKane and T. Newman, {\em Phys. Rev. Lett.} {\bf 94}, 218102 (2005).
\bibitem{mangel} M. Mangel, {\em The theoretical biologist's toolbox} (Cambridge, New York, 2006). 
\bibitem{Fryxell} J. Fryxell, A. Mosser, A. Sinclair, and C.  Packer,  {\em Nature}, {\bf 449}, 1041 (2007).
\bibitem{kaldor} M. Kaldor, {\em New and Old Wars} (Stanford University Press, 1999).
\bibitem{robb} J. Robb, {\em Brave New War} (Wiley, New York, 2007).
\bibitem{ez} V. Eguiluz and M. Zimmermann, {\em Phys. Rev. Lett.} {\bf 85}, 5659 (2000).
\bibitem{neilfin} N. Johnson, P. Jefferies and P. Hui, {\em Financial Market
Complexity}  (Oxford University Press, 2003).
\bibitem{njohnson} N. Johnson, M. Spagat, J. Restrepo, J.
Bohorquez, N. Suarez, E. Restrepo and R. Zarama, e-print
http://lanl.arxiv.org/abs/physics/0605035

\end{thebibliography}
\end{document}